\documentclass[final,5p,times,twocolumn]{elsarticle}

\usepackage{amsmath,amssymb,amsfonts}
\usepackage{graphicx}
\usepackage{textcomp}
\usepackage{xcolor}
\def\BibTeX{{\rm B\kern-.05em{\sc i\kern-.025em b}\kern-.08em
    T\kern-.1667em\lower.7ex\hbox{E}\kern-.125emX}}

\usepackage{comment}
\newcommand{\lmttfont}{\fontfamily{lmtt}\selectfont}

\usepackage{algorithm}
\usepackage[noend]{algpseudocode}

\usepackage{array}
\usepackage{booktabs} 
\usepackage{multirow}

\usepackage{numprint}
\usepackage{nameref}
\usepackage{url}
\usepackage{hyperref}
\usepackage{amsmath}
\usepackage{amsfonts}
\usepackage{amssymb}
\usepackage{xcolor}

\usepackage{color}
\definecolor{comments}{rgb}{0.13,0.55,0.13}
\definecolor{background}{rgb}{0.94, 0.97, 1.0}
\definecolor{strings}{rgb}{0.63,0.125,0.94}
\usepackage{listings}

\makeatletter
\def\lst@makecaption{%
  \def\@captype{table}%
  \@makecaption
}

\usepackage{tikz}
\newcommand*\circled[1]{\tikz[baseline=(char.base)]{
            \node[shape=circle,draw,inner sep=1pt,font=\sffamily\footnotesize] (char) {#1};}}

\journal{The Journal of Systems and Software}

\begin{document}

\begin{frontmatter}



\title{Run-time Failure Detection via Non-intrusive Event Analysis in a Large-Scale Cloud Computing Platform}


\author{Domenico Cotroneo}
\ead{cotroneo@unina.it}
\author{Luigi De Simone}
\ead{luigi.desimone@unina.it}
\author{Pietro Liguori}
\ead{pietro.liguori@unina.it}
\author{Roberto Natella}
\ead{roberto.natella@unina.it}

\address{Università degli Studi di Napoli Federico II}
\address{Naples, Italy}


\begin{abstract}
Cloud computing systems fail in complex and unforeseen ways due to unexpected combinations of events and interactions among hardware and software components.
These failures are especially problematic when they are silent, i.e., not accompanied by any explicit failure notification, hindering the timely detection and recovery.

In this work, we propose an approach to run-time failure detection tailored for monitoring multi-tenant and concurrent cloud computing systems. The approach uses a non-intrusive form of event tracing, without manual changes to the system's internals to propagate session identifiers (IDs), and builds a set of lightweight monitoring rules from fault-free executions. 
We evaluated the effectiveness of the approach in detecting failures in the context of the OpenStack cloud computing platform, a complex and ``off-the-shelf'' distributed system, by executing a campaign of fault injection experiments in a multi-tenant scenario.
Our experiments show that the approach detects the failure with an $F_1$ score ($0.85$) and accuracy ($0.77$) higher than the ones provided by the OpenStack failure logging mechanisms ($0.53$ and $0.50$) and two non--session-aware run-time verification approaches (both lower than $0.15$). 
Moreover, the approach significantly decreases the average time to detect failures at run-time ($\sim 114$ seconds) compared to the OpenStack logging mechanisms.
\end{abstract}




\end{frontmatter}











\section{Introduction}
\label{rv:sec:introduction}
As computer systems grow increasingly complex, they become more error-prone, stemming from their requirements specification, their design, their implementation, or their operating environment \cite{ammar2000comparative}. This is the case of \textit{cloud computing systems}.
Indeed, they include processes distributed across a data center, which often cooperate by message passing and remote procedure calls (e.g., message queues and REST API calls mechanisms). They are quite complex, as they typically consist of software components of millions of lines of code (LoC) that run across dozens of computing nodes.


It is well established that cloud computing systems failures have a huge economic impact on both providers and their customers \cite{li2018empirical,musavi2016experience,gunawi2014bugs,gunawi2016does}. To make things worse, cloud computing systems also fail in complex and unexpected ways. For instance, recent outages reports showed that failures escape fault-tolerance mechanisms, due to unexpected combinations of events and of interactions among hardware and software components, which were not contemplated ahead during the design phase \cite{garraghan2018emergent,hole2019software}. 
These failures are especially problematic when they are \emph{silent}, that is, they are not accompanied by any explicit notification, such as API error codes or error entries in the logs. This behavior hinders timely detection and recovery, makes failures silently propagate through the system, and makes root cause analysis tricky and recovery actions more costly (e.g., reverting the database state) \cite{cotroneo2019enhancing,cotroneo2019bad}. 

Therefore, the prompt detection of the failure at run-time (i.e., \emph{run-time failure detection}) is a key step to improving the fault-tolerance and recovery mechanisms within cloud infrastructures. Generally, logging mechanisms are the main source of information to monitor operation behavior~\cite{farshchi2015experience}, but include several limitations since logs are noisy and lack information on changes in resource states~\cite{oliner2012advances}. 
An effective solution is represented by \emph{run-time verification} strategies, which perform checks over events in the system (e.g., after-service API calls) to assert whether the resources are in a valid state~\cite{bartocci2018introduction}. These checks can be specified as \textit{monitoring rules} using temporal logic and synthesized in a run-time monitor \cite{delgado2004taxonomy,chen2007mop,zhou2014runtime,rabiser2017comparison, cotroneo2018run}. 

Unfortunately, the application of these strategies in the context of cloud computing systems is very challenging~\cite{zhou2014runtime,farshchi2015experience}. In practice, in multi-tenant and concurrent systems, the monitoring rules can be applied as long as the checked events are accompanied by \emph{session identifiers} (IDs). Such IDs allow monitoring solutions to correlate events that belong to the same session (i.e., a set of operations performed on behalf of the same tenant, or by the same subsystem) and to perform checks, e.g., to detect omissions or out-of-order events~\cite{van2009continuous,las2019sifter,horovitz2019non,krause2021design,li2022enjoy}. 
Keeping track of IDs in distributed tracing systems requires intrusive modifications of systems' internals since IDs need to be propagated across API call chains over several components. However, the code instrumentation of the system requires an in-depth knowledge of its internals and is unfeasible for complex and ``off-the-shelf'' systems \cite{parker2020distributed,logz2022introduction}. 
Moreover, this problem is exacerbated by the high number of requests and tenants, which trigger multiple sub-requests within the distributed system. For example, a simple Google search request triggers more than 200 subrequests and crosses hundreds of servers~\cite{zhou2014runtime}. Since requests performed by concurrent tenants may overlap over time, the correlation of events to the same session is a cumbersome task without using any ID. 

Last but not least, events in complex distributed systems are often asynchronous and non-deterministic, thus run-time verification approaches may heavily suffer from false positives/negatives~\cite{satyanarayanan1992transparent,perrochon1998real,cotroneo2019enhancing,cotroneo2020fault}. Sophisticated approaches check formal specifications over events and outputs, by using finite state machines \cite{deligiannis2016uncovering}, temporal logic predicates \cite{arlat2002dependability}, relational logic \cite{gunawi2011fate}, and special-purpose languages \cite{reynolds2006pip}. 
Since these specifications are mostly based on prior knowledge and experience of system designers about failures, they are not meant for discovering new, unknown failure modes of a distributed system, which are missed by the failure specifications. Moreover, writing failure specifications is a time-consuming and cumbersome task, which makes these approaches less applicable in practice. 

To overcome these limitations, in this work we propose an approach (\textit{Monitoring Rules}, MR) to run-time verification tailored for the monitoring and analysis of cloud computing systems. 
The approach uses a non-intrusive form of event tracing that does not require manual changes to the system's internals for propagating IDs. Instead, it automatically analyzes the raw (i.e., unmodified) events already produced by the system (e.g., raw RPC calls and messages over queues); then, it mines relationships among attributes within these events to correlate them; finally, the approach builds a set of lightweight monitoring rules on correlated events from a limited set of ``normal'' (i.e., \textit{fault-free}) executions of the system. 
These rules encode the expected behavior of the system and detect a failure if a violation occurs. 
The proposed approach does not require any in-depth knowledge about the internals of the system, and it is designed to fit in concurrent and multi-tenant environments.

We evaluated the effectiveness of the approach in detecting failures in the context of the OpenStack cloud computing platform by executing a campaign of fault injection experiments in multi-tenant scenarios. OpenStack is a complex and ``off-the-shelf'' distributed system, and represents an important case study as it is one of the most widely deployed open-source cloud software in the world. The complexity of this system, which consists of over 1 million lines of Python code, makes this cloud platform widely targeted by research studies~\cite{mariani2020predicting,ou2018cloud,wu2020cve,cotroneo2021enhancing,zheng2019towards}.
Our experiments show that the approach infers a set of monitoring rules by analyzing $50$ fault-free executions of the system, and detects the failures with an $F_1$ score ($0.85$) and accuracy ($0.77$) higher than the ones provided by the OpenStack failure logging mechanisms ($0.53$ and $0.50$) and two non--session-aware run-time verification approaches (both lower than $0.15$).
Moreover, the approach significantly decreases the average time to detect failures at run-time ($\sim 114$ seconds) compared to the system's failure logging mechanisms.

In the following, Section~\ref{rv:sec:related} discusses related work; 
Section~\ref{rv:sec:approach} presents the overview of the proposed approach;
Section~\ref{rv:sec:openstack} introduces the OpenStack case study;
Section~\ref{rv:sec:rules} describes the process to infer the monitoring rules;
Section~\ref{rv:sec:implementation} shows an implementation of the monitoring rules in a specification language;
Section~\ref{rv:sec:evaluation} experimentally evaluates the approach; 
Section~\ref{rv:sec:threats} discusses the threats to validity; 
Section~\ref{rv:sec:conclusion} concludes the paper. 

\section{Related Work}
\label{rv:sec:related}
In literature, some studies refer to run-time verification as \textit{run-time monitoring} or \textit{dynamic analysis}. 

\vspace{2pt}
\noindent
\textbf{Monitoring and debugging distributed systems.}
Over the last decades, several efforts have been spent on methodologies and tools for monitoring and debugging distributed systems. 
For example, Magpie~\cite{barham2003magpie}, Pinpoint ~\cite{chen2004path}, and \textit{Aguilera et al.}~\cite{aguilera2003performance} identify causal paths in the distributed system, by tracing and correlating call requests and responses, and events at both the OS level and the application server level. 
These approaches were still too difficult to apply in practice, as they either focused only on synchronous (RPC-style) interactions between components and neglected asynchronous and concurrent ones; or, they required intrusive instrumentation of the entire software stack down to the OS. 
Pensieve~\cite{zhang2017pensieve} is a tool for automatically reproducing failures from production distributed systems. Given log files output by the failure execution, the system’s bytecode, a list of supported user commands, and a description of the symptoms associated with the failure, the tool outputs a sequence of user commands, packaged as a unit test, that can reliably reproduce the failure.
Friday~\cite{geels2007friday} is a distributed debugger that allows developers to replay a failed execution of a distributed system, and to inspect the execution through breakpoints, watchpoints, single-stepping, etc., at the global-state level. 
ShizViz~\cite{beschastnikh2016debugging} is an interactive tool for visualizing execution traces of distributed systems, which allows developers to intuitively explore the traces and perform searches; moreover, the tool provides support for comparing distributed executions with a pairwise comparison.
\textit{Gu at al.}~\cite{gu2018kerep} proposed a methodology to extract knowledge about the behavior of the distributed system without source code or prior knowledge. The authors construct the distributed system's component architecture in request processing and discover the heartbeat mechanisms of target distributed systems.
\textit{LOUD}~\cite{mariani2018localizing} is an online metric-driven fault localization technique, which analyses the dependencies among anomalous Key Performance Indicators (KPIs) commonly available in software systems at different abstraction levels to pinpoint the faulty resources that are likely responsible for future failures.
Pip~\cite{reynolds2006pip} is a system for automatically checking the behavior of a distributed system against programmer-written expectations about the system. Pip provides a domain-specific expectations language for writing declarative descriptions of the expected behavior of large distributed systems and relies on user-written annotations of the source code of the system to gather events and propagate path identifiers across chains of requests.
Similar to Pip, Watchtower~\cite{alpernas2021cloud} is a run-time verification tool that analyzes serverless application logs to detect property violations. This tool accepts as input one or more safety properties, and then monitors and analyzes the application at run-time to detect violations.
Cotroneo \textit{et al.}~\cite{cotroneo2019enhancing} proposed a probabilistic approach, based on Variable-order Markov Model~\cite{begleiter2004prediction}, to identify failures in cloud computing systems. The approach showed to be very effective in reducing false positives in a single-user workload.
An \textit{et al.}~\cite{an2017behavioral} demonstrated that system call-based behavioral anomaly detection algorithms can effectively detect previously unknown malware on home routers, which is an essential component to the IoT, with high accuracy and low or no false alarms.

\vspace{2pt}
\noindent
\textbf{Property checking.}
Research studies on run-time verification focused on formalisms for describing properties to be verified. Typically, a run-time verification system provides a \textit{Domain Specification Language} (DSL) for the description of properties to be verified. The DSL can be a standalone language or embedded in an existing language. Specification languages for run-time verification can be regular, which includes temporal logic, regular expressions, and state machines, but also nonregular, which includes rule systems, and stream languages. 
In the run-time verification literature, there is an established set of approaches for the specification of temporal properties, including \textit{Event Processing Language} (EPL). This language is used to translate event patterns into queries that trigger event listeners and determine whether the pattern is observed in an event stream of a \textit{Complex Event Processing} (CEP) environment~\cite{wu2006high}, i.e., a technology for the collection, aggregation, and analysis of sequences of events originated from different sources and occurred at different time. 
Lola~\cite{d2005lola} is a tool implementing run-time verification as a stream computation, where output streams are defined in terms of input streams and/or other output streams. In particular, Lola defines a specification language and algorithms for both online and offline monitoring of synchronous systems and can be used to describe correctness/failure assertions but also statistical measures.
\textit{Zhou et al.}~\cite{zhou2014runtime} proposed a framework that brings run-time verification into the field of trace-oriented monitoring in cloud systems. The monitoring requirements of cloud systems can be specified by formal specification languages, such as finite-state machines, linear temporal logic, etc. 
Atlidakis \textit{et al.}~\cite{atlidakis2020checking} introduced security rules to capture desirable properties of REST APIs and services, and showed how a stateful REST API fuzzer can be extended with active property checkers that automatically test and detect violations of these rules in cloud systems.
\textit{Complex Patterns of Failure} (CPoF)~\cite{power2019providing} is an approach that provides reactive and proactive fault tolerance through complex event processing and machine learning for IoT. The approach uses error events to train ML models to prevent and recover from errors in the future.

\vspace{2pt}
\noindent
Our work exploits run-time verification to state the correctness of a system execution according to specific properties and proposes an approach presenting several points of novelty compared to state-of-the-art studies and tools in run-time verification literature. 
While previous work uses intrusive tracing of the system under test, or requires the check of formal specifications over events or system's outputs, the run-time verification approach proposed in this work i) simply relies on \textit{black-box tracing} to collect the events exchanged in the system (i.e., it does not require knowledge about the system's internals) and ii) models the desired behavior of the system with a set of monitoring rules by analyzing the events collected during the system's operation. These monitoring rules fit cloud computing systems, where we need to face challenging aspects such as multi-tenancy and complex communication flow among the nodes of the system. 

\section{Approach}
\label{rv:sec:approach}
\begin{figure}[t]
    \centering
    \includegraphics[width=1\columnwidth]{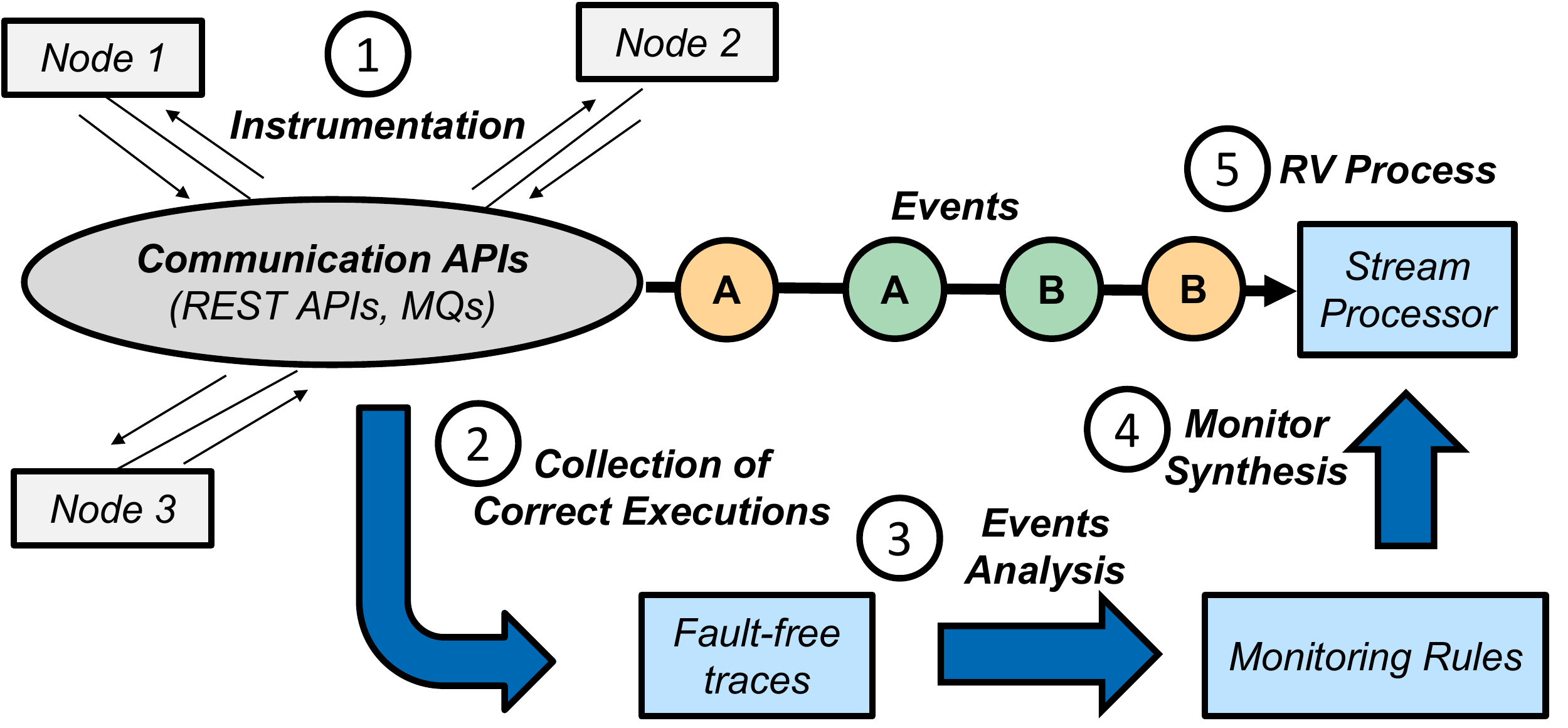}
    \caption{Overview of the proposed approach.}
    \label{rv:fig:approach_overview}
\end{figure}

Figure \ref{rv:fig:approach_overview} shows an overview of the proposed approach. 
The approach is applied to distributed systems within several nodes, each of them providing services that can be requested by message-passing mechanisms.

\begin{figure*}[t]
    \centering
    \includegraphics[width=2\columnwidth]{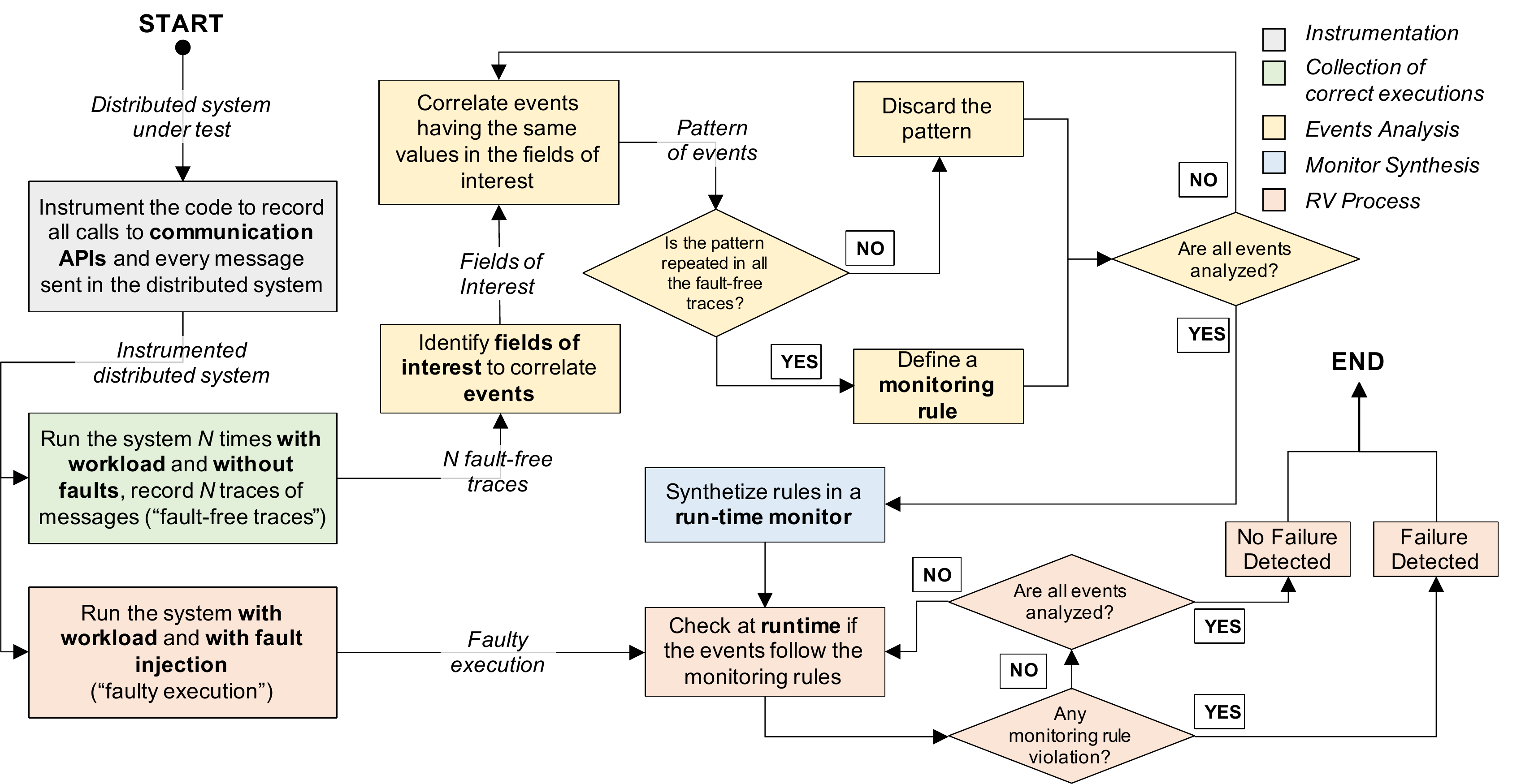}
    \caption{Detailed workflow of the proposed approach}
    \label{fig:flowchart}
\end{figure*}

First, the approach wraps the \emph{communication APIs} of the system, which is instrumented accordingly to collect all messages exchanged by the nodes during operation (step \circled{1}). This instrumentation is a form of ``black-box tracing'' since we collect the messages exchanged among unmodified multi-module system~\cite{koskinen2008borderpatrol} and is especially suitable for complex and distributed systems since it does not require any knowledge about the internals, but only basic information about the communication APIs being used. Moreover, this kind of tracing is already familiar to developers for debugging, performance monitoring and optimization, root cause analysis, and service dependency analysis~\cite{chow2014mystery,chen2002pinpoint}. 
The information recorded by the instrumented APIs includes the time at which a communication API has been called and its duration, the node that invoked the API (\emph{message sender}), and the remote service that has been requested through the API call (\emph{service API}). Moreover, we record information about the response message (e.g., the status code and the message body in an HTTP response, the body of the message, etc.). 
We refer to the calls to communication APIs (i.e., the messages collected during the experiments) as \textbf{\textit{events}}. 
Thus, the system execution produces a \textbf{\textit{trace}} of events that are ordered with respect to the timestamp given by an event collector.
During the system execution, different events can be generated by different calls to the same API service invoked by the same message sender. In this case, we say that the events are of the same \textbf{\textit{type}}.

In step \circled{2}, we collect the \textit{correct executions} of the system. To define its normal (i.e., correct) behavior, we exercise the system under ``fault-free'' conditions, that is, without injecting any faults. Moreover, to take into account the variability of the system, we execute the system $100$ times, collecting different ``\textit{fault-free traces}'', one per execution.

Step \circled{3} analyzes the collected fault-free traces to define a set of \emph{monitoring rules}. These rules encode the expected, correct behavior of the system, and detect a failure if a violation occurs. This step consists of two main operations. The first is selecting key attributes of collected events (e.g., message sender, service API, event timestamp, etc.). Second, we define the failure monitoring rules by inferring \textit{patterns} of events from the fault-free traces. We define a \textit{pattern} as a recurring sequence of (not necessarily consecutive) events, repeated in every fault-free trace and associated with an operation triggered by a workload.
Since we avoid using session identifiers to identify patterns, the approach uses an algorithm based on statistical analysis techniques~\cite{ernst2007daikon,yabandeh2011finding,grant2018inferring} that analyzes the information contained in the body of the events and finds \textit{fields of interest} to correlate different events.

Finally, in step \circled{4} we synthesize an EPL-based \textit{monitor} according to the obtained monitoring rules (Section~\ref{rv:sec:implementation}). 
Because a system failure may cause missing or out-of-order events in the patterns, the monitor processes the stream of events during operation, and it checks, at run-time, whether the system behavior follows the desired behavior specified in the monitoring rules (step \circled{5}). 
To accelerate the occurrence of the failures, the approach performs the \textit{fault-injection experiments}. We focus on injecting one fault per experiment, as injecting multiple faults concurrently is still an open research problem and has not yet been adopted in real projects, due to the high number of combinations among multiple faults.
This step produces \emph{fault-injected traces} (also \emph{faulty traces}), one per experiment. 
Any (run-time) violation of the monitoring rules during the fault-injection experiments alerts the system operator that a failure occurred.

Figure \ref{fig:flowchart} shows a detailed flowchart of the proposed approach.

\section{OpenStack Case Study}
\label{rv:sec:openstack}
OpenStack is a cloud computing platform developed in Python language and is mostly deployed as infrastructure-as-a-service (IaaS) in both public and private clouds where virtual servers and other resources are made available to tenants. 
It provides abstractions and APIs to programmatically create, destroy, and snapshot/revert virtual machine instances; attach and detach volumes and IP addresses; configure security, network topology, and load balancer settings; and many other services to cloud infrastructure consumers. 

OpenStack consists of several independent parts, named \textit{projects} (also referred to as \textit{subsystems}). These projects are developed independently by dedicated teams~\cite{pikeMetric1,pikeMetric2}, each representing a complex distributed system. The three most important subsystems of OpenStack~\cite{denton2015learning,solberg2017openstack} are: (i) Nova, which provides services for provisioning instances (VMs) and handling their life cycle; (ii) Cinder, which provides services for managing block storage for virtual instances; and (iii) Neutron, which provides services for provisioning virtual networks, including resources such as \emph{floating IPs}, \emph{ports} and \emph{subnets} for instances. In turn, these subsystems include several distributed components (e.g., Nova includes \emph{nova-api}, \emph{nova-compute}, etc.), which interact through two communication protocols, i.e., HTTP-based RESTful APIs and remote procedure calls (RPC)~\cite{bahl2012advancing,petrillo2016rest}. 

In OpenStack, tenants can send requests to a service via the dashboard or command line by using the API provided by a specific client developed within each project (e.g., \textit{novaclient} is a client for the OpenStack Compute API).  
The OpenStack API is implemented as a set of web services in the Representational State Transfer (REST) architectural style. An interaction with one of the services involves sending an HTTP-based request to a particular node in the OpenStack cluster and then parsing the response. In the request, we can discern information such as the method invoked (e.g., GET, DELETE, POST, PUSH, etc.), the client performing the request (e.g., \textit{cinderclient}, \textit{neutronclient}, \textit{novaclient}, etc.), and the status code (e.g., 2xx for successful requests, 4xx for client errors, 5xx for server-side errors, etc.).
In the case of the REST API, we identify an event type with the pair client performing the request and the method invoked (e.g., {\lmttfont <novaclient, GET>}).

Furthermore, OpenStack internal subsystems (e.g., \emph{nova-compute}, \emph{cinder-volume}, etc.) use Advanced Message Queuing Protocol (AMQP), an open standard for messaging middleware. This messaging middleware enables the OpenStack services that run on multiple nodes to talk to each other via RPC to serve tenants' requests. 
For example, when a tenant aims to create an instance, it invokes a REST API (i.e., the {\lmttfont /servers} POST method~\cite{openstack_compute_api_server_create}). The request is handled by the Nova subsystem, which starts communicating internally with other subsystems by using remote procedure calls. The first method invoked in the resulting flow of RPC messages is {\lmttfont schedule\_and\_build\_instances}, then Nova exchanges messages with Keystone to verify the tenant's authentication, Glance to get the image, Neutron to create virtual networks, and Cinder for the block storage handling~\cite{instance_image}). 
The RPC messages contain information such as the method invoked, the caller (the system's service), and the body of the message.
In the case of the RPC calls, we identify an event type with the pair subsystem providing the API and method invoked (e.g., {\lmttfont <cinder-volume, create\_volume>}).

\section{Events Analysis}
\label{rv:sec:rules}
We can express a generic monitoring rule by observing the events in the traces.
For example, suppose there is an event of a specific type, say \textit{A}, that occurs before an event of a different type, say \textit{B}, in the same tenant session (i.e., same ID). 
This monitoring rule can be translated into the following pseudo-formalism:

\begin{equation}
     a \to b\ and  \ id(a) = id(b), \ \ with \ a \in A, \  b \in B
\end{equation}

Generally, monitoring rules can be applied in multi-tenant cloud scenarios as long as the information on the tenant IDs is available. However, introducing IDs in distributed tracing systems requires both in-depth knowledge of the internals and intrusive instrumentation of the system. 
Therefore, to make our run-time verification approach easier to apply, we propose a set of coarse-grained monitoring rules (also known as \textit{lightweight monitoring rules}) that do not require the use of any ID. 
To apply the rules in a multi-tenant scenario, we define two different sets of events, A and B, where A and B contain a set of $n$ distinct events of type A and B, respectively, in a time window $[t_0,\ t_0 + \Delta t ]$, assuming $\vert A \vert = \vert B \vert = n$. 


Our monitoring rule for the multi-tenant case then asserts that there should exist a binary relation $R$ over $A$ and $B$ such that:

\begin{equation}
\label{eq:monitoring}
\begin{split}
    R = \{ (a,b) \in A \times B ~|~ & a \to b,\\
    \not\exists ~ a_i, a_j \in A, ~ b_k \in B ~|~ (a_i, b_k), (a_j, b_k), \\
    \not\exists ~ b_i, b_j \in B, ~ a_k \in A ~|~ (a_k, b_i), (a_k, b_j)  ~ \}\\
\end{split}
\end{equation}

\noindent
with $i,j,k \in [1,n]$. 
That is, every event in $A$ has an event in $B$ that follows it, and every event $a$ is paired with exactly one event $b$, and vice-versa. 
These rules are based on the observation that, if a group of tenants performs concurrent operations on shared cloud infrastructure, then a specific number of events of type A is eventually followed by the same number of events of type B. The idea is inspired by the concept of flow conservation in network flow problems. Without using a propagation ID, it is impossible to verify the happened-before relation between the events \(a_i\) and \(b_i\) that refer to the same session or the same tenant \(i\), but it is possible to verify that the total number of events of type A is equal to the total number of events of type B in a pre-defined time window.

\subsection{Events Correlation}
\label{rv:subsec:algorithm}

To specify a monitoring rule, we need to identify one or more events characterizing the action taken by the tenant. 
In the example described in Section~\ref{rv:sec:openstack}, the first RPC message exchanged among the subsystems includes the invocation of method {\lmttfont schedule\_and\_build\_instances} by the {\lmttfont nova-conductor} component (it provides coordination and database query support for Nova). This RPC event follows the {\lmttfont /servers} POST method called by \textit{novaclient}, but not every {\lmttfont <novaclient, POST>} event generates the {\lmttfont schedule\_and\_build\_instances} call since the POST method can be used to create/add different resources, i.e., there is not a one-to-one relationship between the first RPC message of the event flows and the REST API starting the request. 

Because we cannot discern patterns from the observation of the REST API calls, we need to look at the RPC messages.
If we observe the method {\lmttfont schedule\_and\_build\_instances} invoked by {\lmttfont nova-conductor} component, we infer that the tenant requested the creation of an instance. Similarly, if we observe the method {\lmttfont create\_volume} invoked by the {\lmttfont cinder-scheduler} component (used to determine how to dispatch block storage requests), then we derive that the tenant aims to create a volume, and so on. By taking this into account, we refer to the first (with the lower timestamp) event that occurred in the pattern of RPC events as \textit{\textbf{head event}}.

To identify the monitoring rules, the approach focuses on finding patterns of events starting with a head event. The key idea is that, if we find a pattern of recurring events starting from a specific head event, then we can specify the rules to identify anomalies (e.g., out-of-order events, missing events, etc.).
Unfortunately, due to the non-determinism of cloud systems, we can not manually infer rules by simply observing fault-free executions. Indeed, the head event starting from a tenant request is not necessarily followed by the same number and/or the same order of events. Moreover, the high volume of messages in the system makes manual inspection very difficult and prone to errors.


To correlate the events in a pattern without using any session IDs, the approach analyzes the \textit{fields} in the \textit{body} of the RPC messages.
The key idea is that, even if there is no common field to all events in a session or a trace, a subset of the sub-requests in the request flow may have some fields in common, such as the ID of a virtual resource (e.g., a volume, an instance), the tenant name, etc.
Since the manual analysis of the body of the RPCs requires domain-specific knowledge of the system internals~\cite{sharma2015hansel}, the approach keeps this operation lightweight by using an algorithm that analyzes all the events within a set of fault-free traces.
More specifically, the algorithm analyzes all the fields $f$ in the body of the RPC events collected in a set of fault-free traces and returns only the fields (\textit{fields of interest}) that satisfy the following properties:

\vspace{0.1cm}
\noindent
$\blacksquare$ \textit{\textbf{$P_1$}:} \textit{In every fault-free trace, the propagation of the values assumed by a filed $f$ should be higher than a threshold, say $\epsilon_1$}. This property expresses that, in order to correlate events with a generic field $f$, then the values assumed by the field should ideally propagate across the fields of different events in the fault-free trace. 

\vspace{0.1cm}
\noindent
$\blacksquare$ \textit{\textbf{$P_2$}:} \textit{In every fault-free trace, the number of non-unique values assumed by a field $f$ should be higher than a threshold, say $\epsilon_2$}. Indeed, if a value assumed by a generic field $f$ is repeated across all the events, then the property \textit{$P_1$} would correlate all the events of the trace in a single pattern. Therefore, this property ensures that the field $f$ assumes different values across the uncorrelated events.

\begin{figure}
    \centering
    \includegraphics[width=1\columnwidth]{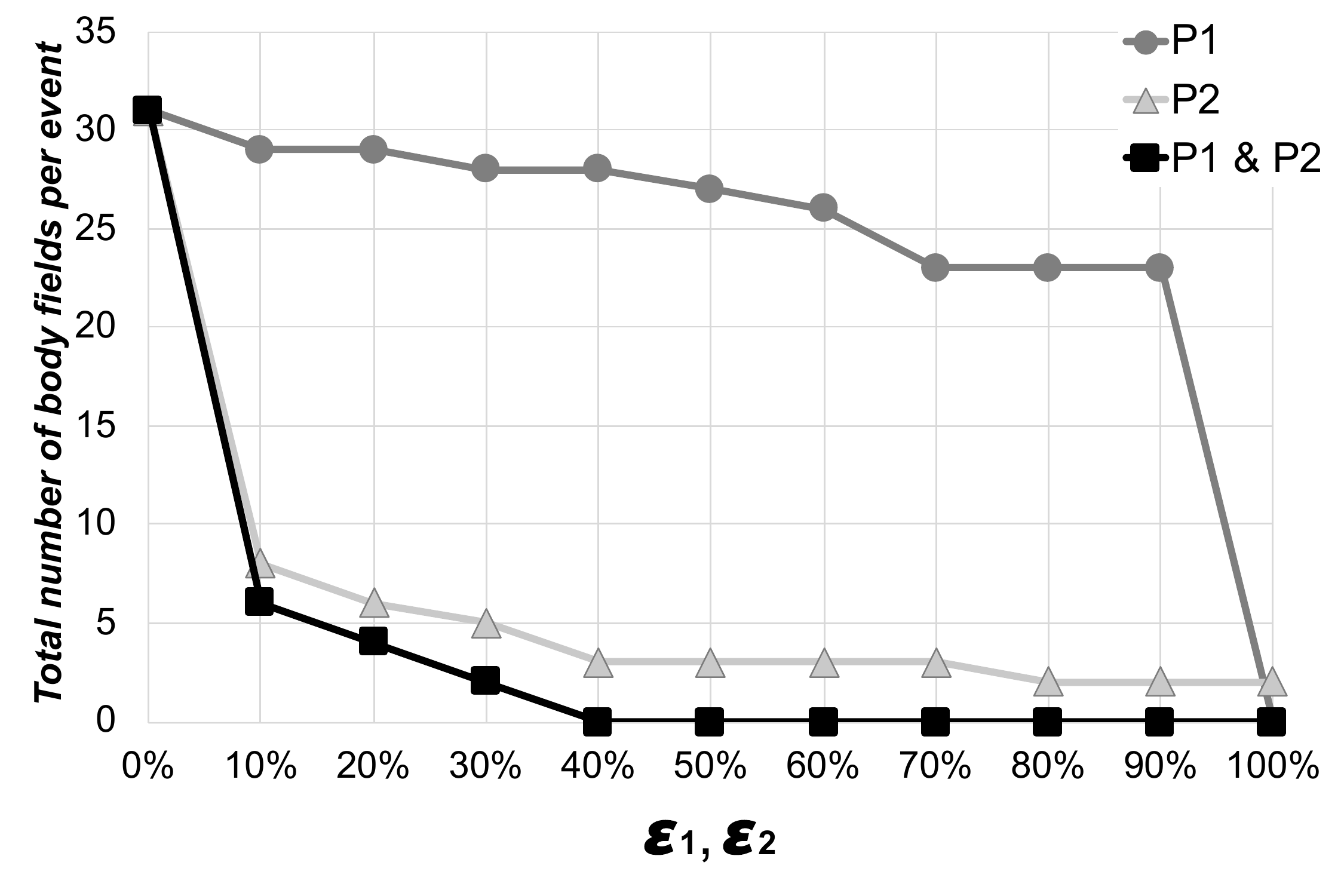}
    \caption{Sensitivity analysis of the thresholds $\epsilon_1$ and $\epsilon_2$}
    \label{fig:epsilon}
\end{figure}

To assess how the choice of the thresholds $\epsilon_1$ and $\epsilon_2$ impacts the number of discarded fields in the body of the RPCs, we performed a sensitivity analysis by varying the thresholds between $0\%$ and $100\%$, as shown in \figurename{}~\ref{fig:epsilon}. 
The analysis highlights that the property $P_1$ is always satisfied also when the threshold $\epsilon_1$ is high. In particular, almost the total number of the fields $f$ propagate their values across the $90\%$ of remaining events in the trace.
However, most of the propagation of these values is due to fields that always assume a single value across almost all the events in a trace. This is the case for fields such as the OpenStack project ID (i.e., organizational units in the cloud to which administrators can assign tenants~\cite{openstack_project}), which assume the same values across all the events in the system executions. Indeed, the figure shows that only less than one-third of the total number of the fields $f$ assumes non-unique values for at least the $10\%$ of times in the trace ($P_2$).
Combining both properties, we obtain that all the fields are discarded when both thresholds are $ \geq 40\%$. 
Therefore, to limit the number of fields of interest, we made a conservative choice of both thresholds, by setting them equal to $30\%$.

Once defined the properties, the algorithm filters the fields that do not satisfy both of them, working accordingly to the following steps: 

\begin{enumerate}
    \item Define the empty sets $F = \{\}$, $\Phi = \{\}$;
    
    \item Extract all fields $[f_1, f_2, ..., f_N]$ from the body of all events in the traces and add them to the set $F$;
    
    
    
    \item For every field $f_i$, with $i \in [1,N]$, if the field does not satisfy both properties $P_1$ and $P_2$, then $f_i \cup \Phi$;
    
    
    \item Return $F \setminus \Phi$.
    
\end{enumerate}

The application of the algorithm massively reduces the number of fields that can be analyzed manually. In our case study, from many fields in the body of the RPC messages (see Section~\ref{subsec:setup}), the algorithm returns the parameters used in the {\lmttfont oslo\_context.context} of \textit{Oslo Context} library, a base class for holding contextual information of a request~\cite{oslo_context}.
More specifically, the algorithm returns the variable {\lmttfont \_context\_request}, i.e., the identifier of a request, and the variable {\lmttfont \_context\_global\_request}, i.e., a request-id sent from another service to indicate that the event is part of a chain of requests~\cite{oslo_context}.
The values of these variables are hexadecimal strings prefixed by ``req-"~\cite{sharma2015hansel}.

The approach correlates the events with the same values in fields {\lmttfont \_context\_request} and {\lmttfont \_context\_global\_request} of the body of the RPC messages.
However, since correlated events may occur too far in time (the workload execution may last tens of minutes or even hours), the approach defines a max time length of the pattern, that is, the temporal distance between the last event and the first event (i.e., the head event) in the chain of correlated events is lower than the length of a \textit{time window} $\Delta T$. \tablename{}~\ref{tab:correlation} shows an example of the correlation process. Each event in the trace has several attributes, such as the timestamp, the sender, the service API, the fields of the body, etc. When, in a specified time window $\Delta T$, the approach finds events having the same values in the field {\lmttfont \_context\_global\_request} or in the field {\lmttfont \_context\_request} of an event that occurred before, then it defines a pattern of events in the fault-free trace. 
In the example of the table, the approach defines a pattern containing the events {\lmttfont <consoleauth, delete tokens for instance>} and {\lmttfont <compute, terminate instance>} because they have the same {\lmttfont \_context\_request}. Then, it extends the pattern with three further events having the {\lmttfont \_context\_global\_request} equal to the {\lmttfont \_context\_request} of the two events in the pattern. The final pattern contains $5$ events. Notice that no more events can be added to this pattern since the pattern length can not be higher than the time window. 

\begin{table*}[t]
    \centering
    \caption{Correlation of the events in a pattern. By looking at the context request field, the approach finds two different patterns of 2 (\textbf{\textcolor{red}{red color}}) and 3 events (\textbf{\textcolor{blue}{blue color}}), respectively. The two patterns are merged into a single one of 5 events since the values of the context request of the first pattern propagate across the context global request field of the second pattern.}
    \label{tab:correlation}
    \includegraphics[width=2\columnwidth]{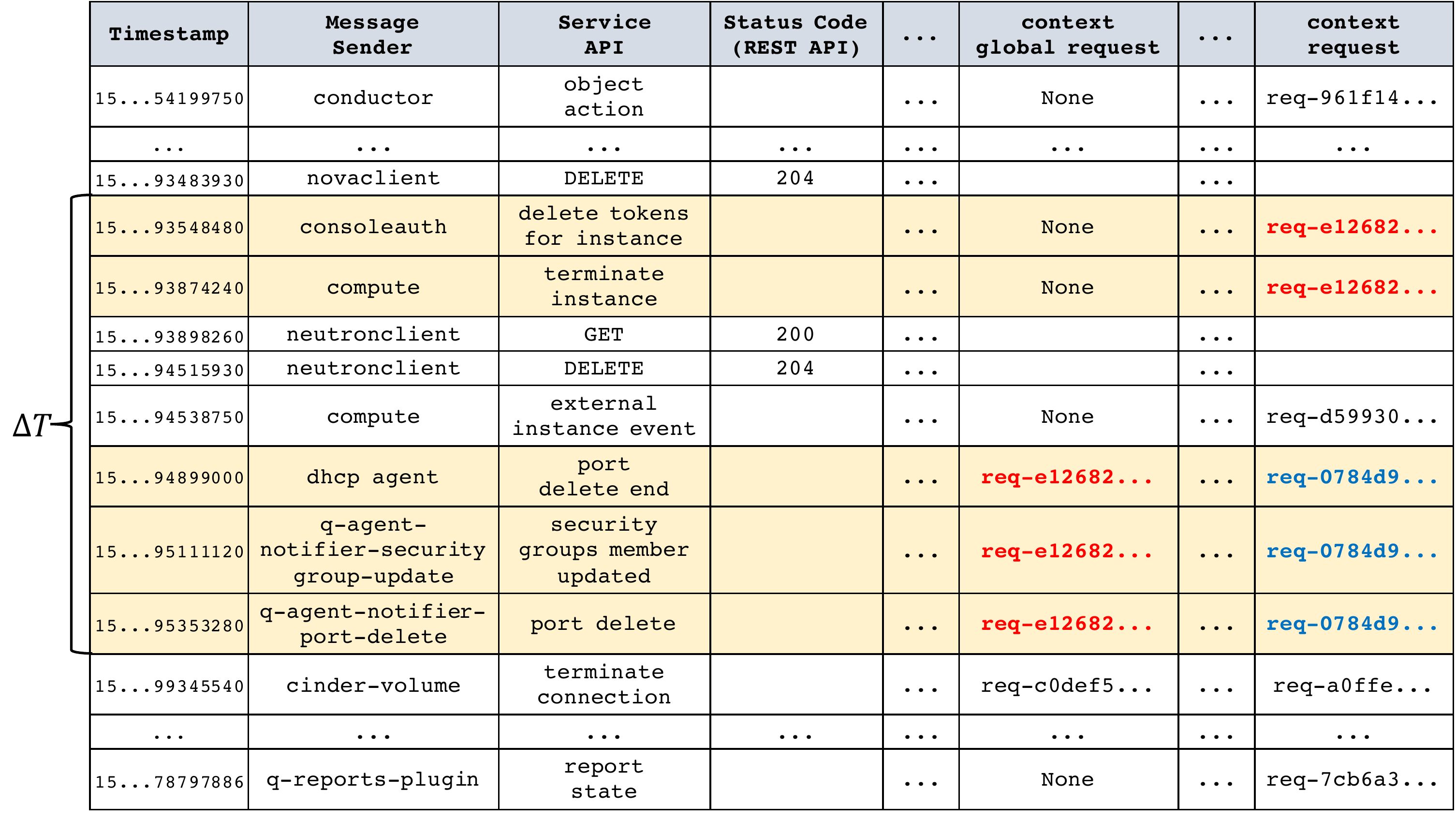}
\end{table*}

When a pattern of - not necessarily consecutive - events is repeated over all the fault-free executions, the approach defines a monitoring rule.
For example, if the $5$ events of the pattern shown in the example of \tablename{}~\ref{tab:correlation} occur (regardless of the order) in all the fault-free executions of the system, then a new rule is added to the set of the monitoring rules.
Therefore, suppose to have $m$ fault-free executions, and consider $p_{A,1}, p_{A,2}, ..., p_{A,m}$ as the patterns starting with the same head event type $A$ found by the approach in the different executions. To define $MR_{A}$, i.e., the monitoring rule activated by the event type $A$, the approach finds all the common events in the different patterns: 

\begin{equation}
    MR_{A} = p_{A,1} \cap p_{A,2} \cap ... \cap p_{A,m} 
\end{equation}

$MR_{A}$ will contain at least one event $a \in A$ because all the patterns start with the same head event.
To check if the system's behavior is the desired one, we need at least one event following the head event used to activate the rule. Therefore, if at least one of the subsequent events is common to all the patterns, i.e., if $|MR_{A}| \geq 2$, then the approach defines a new monitoring rule.

\subsection{Rules Classification}
\label{rv:subsec:classification}
Since the events in a pattern may occur in a different order, we classified the rules according to three different categories, explained in the following.
Suppose to observe, in a specified time window, three different RPC events, say $a, b, c$ belonging to three different event types, say $A, B, C$, respectively, and that the event $a$ is the head event, i.e., the occurrence of this event identifies a pattern of events that follow the heading one. 
We categorize the monitoring rules as follows.

\vspace{2pt}
\noindent
$\blacksquare$ \textbf{\textit{Ordered-Events}} (ORD): Rules based on a flow of events that always follows the same order and occurrence. For example, the event $b$ and $c$ follow $a$ always with the same pattern (e.g., $a \to b \to c$). 
Therefore, when the event $a$ occurs, then the approach waits for the occurrence of the event $b$ and the event $c$, with $c$ following $b$, in a specified time window $\Delta T$.
These rules characterize the services less affected by the non-determinism and where it is possible to find a fixed pattern for the same operation. The ORD rules can detect failures causing out-of-order or missing events during the system execution. 

\vspace{2pt}
\noindent
$\blacksquare$ \textbf{\textit{Occurred-Events}} (OCC): Rules based on a flow of events that occur after the head event without following any specific order and/or number. For example, this happens when the event type $b$ occurs before or after event type $c$ (e.g., $a \to b \to c$ or $a \to c \to b$). 
Therefore, when the event $a$ occurs, then the approach waits for the occurrence of the event $b$ and the event $c$, without a fixed ordering, in a specified time window $\Delta T$.
These rules take into account the non-determinism of the flow of events, i.e., we cannot identify a fixed pattern among all the system executions.
The OCC rules can detect failures causing missing events in the pattern, but not out-of-order events.

\vspace{2pt}
\noindent
$\blacksquare$ \textbf{\textit{Counted-Events}} (COUNT): Rules based on the observation that an event (or more events) is repeated several times, varying in a range of value (e.g., $min_{count} < a < max_{count}$, where $min_{count}$ and $max_{count}$ represent the minimum and the maximum number of times the event is repeated under fault-free conditions, respectively). The COUNT rules can detect failures when the system is unable to serve a request involving multiple-repeated operations, such as polling requests on a resource. In this case, a failure leads to an anomalous repetition of events (i.e., $a > max_{count}$) since the requests are issued multiple times.

\section{Monitor Implementation}
\label{rv:sec:implementation}
After identifying the monitoring rules, we synthesize the rules in a run-time monitor that verifies whether the system’s behavior follows the desired one. Any run-time violation of the monitoring rules gives a timely notification to avoid undesired consequences, e.g., non-logged failures, non-fail-stop behavior, failure propagation across subsystems, etc. 

\subsection{Implementation}
We translate the rules in the \textit{Event Processing Language} (EPL), a particular specification language provided by the \textit{Esper} software~\cite{esper}, and allow the expression of different types of rules (i.e., temporal, statistical, etc.).
The EPL extends the SQL standard language, offering both typical SQL clauses (e.g., {\lmttfont select}, {\lmttfont from}, {\lmttfont where}, {\lmttfont insert into}) and additional clauses for event processing (e.g, {\lmttfont pattern}, {\lmttfont output}). 
The \textit{Esper compiler} compiles EPL source code into Java Virtual Machine (JVM) bytecode so that the resulting executable code runs on a JVM within the \textit{Esper runtime} environment.


We applied the EPL statements derived from the monitoring rules to detect failures in OpenStack when multiple tenants perform requests concurrently. 
Since we do not collect a tenant ID, we use a \textit{counter} to take into account multi-tenancy operations. 
To estimate the number of concurrent requests performed by different tenants, we associate a counter to every monitoring rule and increment its value every time we observe the head event.
For example, if we observe twice the event type {\lmttfont <nova-conductor, schedule\_and\_build\_instances>} (i.e., the head event of the request flow related to the instance creation) in the same time window, then we activate twice the monitoring rule since two different tenants are requesting the creation of an instance. 
The value of the counter is sent, along with the event name, to the \textit{Esper runtime} component.
Listing~\ref{lst:volume_creation} shows the EPL translation of the rule \textit{Volume Creation}.

\noindent
\begin{minipage}{1\columnwidth}
\begin{lstlisting}[basicstyle=\footnotesize, label={lst:volume_creation}, caption={OpenStack Volume Creation rule in EPL}]
@name('VolumeCreation') select * from pattern 
[every a = Event (name = "cinder-scheduler_create_volume") -> 
(timer:interval(secondsToWait seconds) and not b = Event 
(name = "cinder-volume_create_volume", countEvent = a.countEvent))];
\end{lstlisting}
\end{minipage}

When the \textit{Esper runtime} observes the (head) event {\lmttfont <cinder-scheduler, create\_volume>} with its counter value, it waits for the event {\lmttfont <cinder-volume, create\_volume>} with the same counter value in a time window of {\lmttfont secondsToWait} seconds. If this condition is not verified, the approach notifies of a failure. 

To express the monitoring rule, we used the clause {\lmttfont pattern} (to define a pattern of events), 
and the operators {\lmttfont every}, {\lmttfont followed-by} ($\to$), and {\lmttfont timer:interval}. 
The operator {\lmttfont every} defines that every time we observe part of the pattern (e.g., the observation of the head event {\lmttfont cinder-scheduler, create\_volume} in Listing \ref{lst:volume_creation}), the \textit{Esper runtime} actives a monitoring rule. Without this operator, the monitoring rule would be activated only once.
The operator $\to$ defines the order of the events in the rule, while the operator {\lmttfont timer:interval} establishes the length of the time window.

The synthesis of the monitor is automatically performed once EPL rules are compiled. The \textit{Esper runtime} acts like a container for EPL statements, which continuously executes the queries (expressed by the statements) against the flow of events. We invite the reader to refer to the official documentation for more detailed information on Esper~\cite{esperDoc}.


\subsection{Events Collection}
To collect the events exchanged in the system, we adopt the \emph{Zipkin} distributed tracing system \cite{zipkin}, due to its maturity, high performance, and support for several programming languages~\cite{salesforce_blog,twitter_blog}. 
We instrument APIs to send data via HTTP to the \textit{Zipkin collector}, which stores trace data. The collected events are ordered according to a timestamp given by the collector. In this work, we instrumented the following communication points in OpenStack:

\begin{itemize}

    \item The \emph{OSLO Messaging library}, which uses a message queue library to exchange messages with an intermediary queuing server (RabbitMQ) through RPCs. These messages are used for communication among OpenStack subsystems.
    
    \item The \emph{RESTful API libraries} of each OpenStack subsystem, i.e.,  \emph{novaclient} for Nova (implements the OpenStack Compute API \cite{computeAPI}), \emph{neutronclient} for Neutron (implements the OpenStack Network API \cite{networkAPI}), and \emph{cinderclient} for Cinder (implements the OpenStack Block Storage API \cite{storageAPI}). These interfaces are used for communication between OpenStack and its clients (e.g., IaaS customers).
    
\end{itemize}

\textit{Zipkin} puts a negligible overhead in terms of run-time execution since it adopts an asynchronous collection mechanism to avoid critical execution paths. Moreover, we only instrument $5$ selected lines of communication system code (e.g., the {\lmttfont cast} method of OSLO to broadcast messages), by adding simple annotations (the Zipkin context manager/decorator) only at the beginning of these methods (a total of $21$ lines of Python code). Our instrumentation neither modified the internals of OpenStack subsystems nor used any domain knowledge. 

We extract periodically the events stored in the \textit{Zipkin collector} and information such as the invoked method, the service providing the API, the timestamp, the body of the RPC messages, and the status code of the REST API. The processed information is then pushed into a queue, named \textit{Esper Inputs Waiting Queue}, which stores the flow of events. The events in the queue are sent as inputs to the \textit{Esper runtime}, which compares the flow of events against every statement compiled by the \textit{Esper compiler} (i.e., the monitoring rules): if that event satisfies the condition specified in a rule, then the rule moves to the next condition, otherwise, it raises an exception, notifying an unexpected behavior.

\section{Experimental Evaluation}
\label{rv:sec:evaluation}
We evaluate the proposed approach by performing fault injection experiments against the OpenStack cloud management platform, which is a relevant case study since it is a large and complex distributed system.



\subsection{Setup}
\label{subsec:setup}
We targeted OpenStack version 3.12.1 (release \emph{Pike}), deployed on Intel Xeon servers (E5-2630L v3 @ 1.80GHz) with 16 GB RAM, 150 GB of disk storage, and Linux CentOS v7.0, connected through a Gigabit Ethernet LAN. 

To evaluate the approach in realistic scenarios, we developed a multi-tenant workload generator, which simulates $10$ different tenants performing concurrent operations on the cloud infrastructure.
The tenants exhibit $6$ different profiles, as described in the following:
\begin{itemize}

    \item \textit{\textbf{Volume Only}}: The tenant performs operations strictly related to the block storage (Cinder subsystem);
    
    \item \textbf{\textit{Instance Only}}: The tenant stresses the Nova subsystem for the creation of VM instances;
    
    \item \textbf{\textit{Network Only}}: The tenant creates network resources (networks, sub-networks, IP addresses, routers, etc.), stressing the Neutron subsystem;
    
    \item \textbf{\textit{Instance before Volume}}: The tenant creates an instance from an image, then a storage volume;
    
    \item \textbf{\textit{Volume before Instance}}: The tenant creates a volume and then an instance starting from the volume;
    
    \item \textbf{\textit{Instance, Volume, and Network}}: The tenant stresses the Nova, Cinder, and Neutron subsystems in a balanced way. 
    
\end{itemize}

These six profiles are run concurrently to generate a multi-tenant workload. The \textit{Volume Only}, \textit{Network Only}, \textit{Instance before Volume}, and \textit{Volume before Instance} profiles are run twice by different tenants. 

The execution of the workload lasts $\sim40$ minutes and produces a large amount of data. On average, during every fault-free execution, we collected \numprint{89} different event types (\numprint{75} RPC and \numprint{14} REST API), and $\sim$\numprint{2,400} different events ($\sim$\numprint{2,050} RPC events, while the remaining are related to the REST API calls). For every execution trace, the bodies of all RPC events contain in total more than \numprint{60,000} fields (on average, $\sim31$ body fields per RPC event).


\subsection{Fault-free Analysis}

\begin{table}[t]
\centering
\caption{Monitoring Rules in the fault-free scenario.}
\label{rv:tab:monitoring_rules}
\small
\begin{tabular}{   
>{\centering\arraybackslash}m{2cm}
>{\centering\arraybackslash}m{1.5cm} 
>{\centering\arraybackslash}m{1.5cm} 
>{\centering\arraybackslash}m{1.5cm}}
\toprule
\textbf{Rule Description} & \textbf{Rule Type}    & \textbf{\# of Events} & \textbf{Subsystems} \\ \midrule
\textit{Instance Creation}      &   ORD     &   4       &   Nova\\
\textit{Volume Creation}        &   ORD	    &   2	    &   Cinder\\
\textit{Network Creation}	    &   OCC	    &   3	    &   Neutron\\
\textit{Volume Attachment}      &   ORD     &   4       &   Nova, Cinder\\
\textit{Instance Deletion}      &   ORD	    &   3       &   Nova\\
\textit{Security Group Update}  &   ORD     &   2       &   Neutron\\
\textit{Ping Instance via SSH}  &   COUNT	&   6-26    &   Neutron\\
\bottomrule
\end{tabular}
\end{table}

Since the execution of the system execution is not trivial and time-consuming, we limited the number of fault-free traces. Indeed, we collected $50$ fault-free traces, exercising the system with the multi-tenant workload. 
To set the time window $\Delta T$ and specify the length of the patterns, we made a conservative choice by setting it equal to the maximum time needed by OpenStack to serve any request performed by the multi-tenant workload in fault-free conditions ($\sim 35$ seconds, in our testbed). Although this duration depends on both the workload, i.e., the operations performed during the experiments, and the hardware where OpenStack is deployed (high hardware requirements imply a shorter time to serve the requests), it can be easily computed by running the workload in fault-free conditions (e.g., by using the logs of the workload). 

We derived $7$ types of monitoring rules based on RPC messages, as shown in \tablename{}~\ref{rv:tab:monitoring_rules}.
The rules include the creation of resources, such as instances, volumes, and networks, which are common operations on an IaaS cloud.  
The rules related to the creation of the instance and volume are of type ORD, while the one related to the creation of the network is OCC. We attribute this to the asynchronous nature of the Neutron subsystem.
The approach also identified the rule for the attachment of the volume to an instance and the deletion of the instance. Moreover, the approach derived two further rules related to the network operations: the update of the security groups (the sets of network filter rules that are applied to all instances, e.g., allowed/disallowed SSH traffic, etc.) that define networking access to the instance, and the connection to an instance via SSH, which is the only rule of type COUNT. 


We notice that the monitoring rules inferred by our approach do not encompass all possible operations performed by the workload.
Indeed, \textit{volume deletion} and \textit{instance reboot} are notable examples of operations not included in \tablename{}~\ref{rv:tab:monitoring_rules}. 
We investigated the fault-free traces and observed that these operations do not involve a sequence of events, but only a single head event. However, to monitor the system, we need a pattern of at least two events, i.e., at least one event has to follow the head event activating the rule in a temporal window. 

\subsection{Fault Injection Experiments}

We evaluated our approach by performing a fault-injection campaign in OpenStack. In total, we performed $637$ experiments by injecting faults in Nova, Cinder, and Neutron subsystems (one fault per experiment).
To perform the experiments, we developed a tool~\cite{cotroneo2020profipy,cotroneo2019failviz} to automatically scan the source code of OpenStack, find all injectable API calls, and inject faults by mutating the calls. The tool identifies the injectable locations that are actually covered by the running workload and performs one fault injection test per covered location. 
To define the faults to inject into the target system, we analyzed over $179$ problem reports on the OpenStack bug repository. This analysis allowed us to identify the most recurrent bugs in OpenStack over the last few years. In particular, we choose the following faults, which are among the most frequent in OpenStack~\cite{cotroneo2019bad}:

\begin{itemize}
    \item \textbf{\textit{Throw exception}}: An exception is raised on a method call, according to a pre-defined, per-API list of exceptions.
    \item \textbf{\textit{Wrong return value}}: A method returns an incorrect value. The wrong return value is obtained by corrupting the targeted object, depending on the data type (e.g., by replacing an object reference with a null reference, or by replacing an integer value with a negative one).
    \item \textbf{\textit{Wrong parameter value}}: A method is called with an incorrect input parameter. Input parameters are corrupted according to the data type, as for the previous point.
\end{itemize}

Before every experiment, we clean up any potential residual effect from the previous experiment, to be able to relate failure to the specific fault that caused it. We redeploy the cloud management system, remove all temporary files and processes, and restore the OpenStack database to its initial state. 

During the execution of the workload, any exception generated by API calls (\emph{API Errors}) is recorded. In between calls to service APIs, the workload also performs \emph{assertion checks} on the status of virtual resources, to point out failures of the cloud management system. 
These checks assess the connectivity of the instances through SSH and query the OpenStack API to ensure that the status of the instances, volumes, and network is consistent with the expectation of the tests. 
In our context, assertion checks serve as \emph{ground truth} about the occurrence of failures during the experiments. These checks are valuable in identifying the cases where a fault causes an error and the system does not generate an API error (i.e., the system is unaware of the failure state) \cite{cotroneo2019bad}.

We consider an experiment as failed if at least one API call returns an API error or if there is at least one assertion check failure. 
In total, we observed failures in \numprint{496} experiments ($\sim78\%$ of the total number of experiments).  
In the remaining tests, there were neither API errors nor assertion failures since the fault did not affect the behavior of the system (e.g., the corrupted state is not used in the rest of the experiment, or the error was tolerated). This is a typical phenomenon that occurs in fault injection experiments \cite{christmansson1996generation,lanzaro2014empirical}; yet, the experiments provided us with a large and diverse set of failures for our analysis. 

In many failures, when the tenant performs a request by using the REST APIs of the system, the events related to these calls contain a status code $4xx$ or $5xx$, indicating the incapability of the client/server to perform/serve the request. These events cannot be observed during fault-free executions since they reflect failure symptoms.
In these cases, the flow of RPC events starting from the REST API call does not occur, making the RPC events-based rules not effective in detecting anomalies. Therefore, to support the monitoring rules, the approach notifies the failure when we observe a REST API call with status code $4xx$ or $5xx$.

To help the research community in the application and evaluation of new solutions for detecting failures in the system, we shared on GitHub\footnote{\href{https://github.com/dessertlab/OpenStack-multi-tenant-workload}{https://github.com/dessertlab/OpenStack-multi-tenant-workload}} the raw logs collected during the execution of the OpenStack cloud computing platform with the multi-tenant workload, by including both the fault-injection experiments and the fault-free executions of the system. The repository also contains the Python code we used to analyze the execution traces and infer the monitoring rules, and the scripts for the multi-tenant workload in the OpenStack cloud computing platform.

\subsection{Evaluation Metrics}
\label{rv:subsec:evaluation_metric}
We evaluated the approach in terms of \textit{precision} and \textit{recall}. 
The former is mathematically computed as the number of \textit{true positives} identified by the approach over the total number of positives predicted (\textit{true} and \textit{false positives}).
The latter, instead, is computed as the number of \textit{true positives} identified by the approach over the total number of actual positives (\textit{true positives} and \textit{false negatives}).
We consider failure detection as a true positive case only when the approach detects the ``first'' failure of the system. For example, if OpenStack fails to create an instance and the approach detects the failure only on the subsequent attachment of a volume to the failed instance, we consider the experiment as a \textit{false negative} case since the first failure experienced by the system (i.e., the instance creation) was undetected. This conservative choice is due to the need to detect failures as soon as they occur in the system and avoid error propagation. The \textit{false positives} cases, instead, refer to the experiments in which the approach identifies a failure before the actual failure of the system or when the system is not failed at all.
To perform a comprehensive evaluation, we use the \textit{$F_1$ score}, defined as the harmonic mean of the precision and recall. Moreover, to assess also the \textit{false negatives}, we adopted the \textit{accuracy} as a further metric, which is the ratio of correct detections (i.e., both \textit{true positives} and \textit{true negatives}, where the latter refers to the cases in which the approach does not notifies any failure and the system is actually not failed) over the total number of experiments.
All metrics range from $0$ (total misclassification) to $1$ (perfect classification).

\subsection{Experimental Results}
\label{rv:subsec:experimental_results}

\begin{table*}[ht]
\centering
\caption{Approaches comparison. The best performance is \textbf{bold}. The worst performance is \textcolor{red}{\textbf{red/bold}}. Time window $\Delta T = 35 $ s (MR approach).}
\label{rv:tab:evaluation}
\small
\begin{tabular}{ 
>{\centering\arraybackslash}m{2.5cm} |
>{\centering\arraybackslash}m{2.5cm} |
>{\centering\arraybackslash}m{1.7cm}
>{\centering\arraybackslash}m{1.7cm}
>{\centering\arraybackslash}m{1.7cm} 
>{\centering\arraybackslash}m{1.7cm}} 
\toprule
\textbf{OpenStack Subsystem}          & \textbf{Approach}     & \textbf{Precision} & \textbf{Recall} & \textbf{$F_1$ \textbf{score}} &\textbf{Accuracy} \\
\midrule
\multirow{5}{*}{\textit{Nova}}      & \textit{FL}          &   \textbf{1.00}    &  0.30            &   0.46            &  0.47 \\
                                    & \textit{UN}           &   0.26      &  \textbf{\textcolor{red}{0.11}}   &    0.16 & 0.09  \\
                                    & \textit{PM}           &   \textbf{\textcolor{red}{0.05}}            &  0.30    &   \textbf{\textcolor{red}{0.08}}    &   \textbf{\textcolor{red}{0.04}} \\
                                    & \textit{MR}           &   0.89            &  \textbf{1.00}    &   \textbf{0.94}   &   \textbf{0.91} \\
                                    & \textit{FL with MR}  &   0.89            &  \textbf{1.00}    &   \textbf{0.94}   &   \textbf{0.91} \\
                                    \midrule
                                    
\multirow{5}{*}{\textit{Cinder}}    & \textit{FL}          &   \textbf{1.00}   &  0.28             &   0.44   &   0.38 \\
                                    & \textit{UN}           &   \textbf{\textcolor{red}{0.11}}      &  \textbf{\textcolor{red}{0.02}}     &  \textbf{\textcolor{red}{0.03}}   &  \textbf{\textcolor{red}{0.02}} \\
                                    & \textit{PM}           &   0.13            &  0.41    &   0.20   &   0.11 \\
                                    & \textit{MR}           &   0.85    &  0.84    &  \textbf{0.85} &   \textbf{0.74} \\
                                    & \textit{FL with MR}  &   0.85    &  \textbf{0.85}            &   \textbf{0.85}  &   \textbf{0.74} \\
                                    \midrule
                                    
\multirow{5}{*}{\textit{Neutron}}   & \textit{FL}          &   \textbf{1.00}    &  0.71    &   0.83   &   0.80 \\
                                    & \textit{UN}           &   \textbf{\textcolor{red}{0.12}}    &  \textbf{\textcolor{red}{0.06}}     &  \textbf{\textcolor{red}{0.08}}   &   \textbf{\textcolor{red}{0.04}} \\
                                    & \textit{PM}           &   0.13            &  0.32    &   0.18   &   0.10 \\
                                    & \textit{MR}           &   0.87    &  0.31    &   0.46 &  0.50 \\
                                    & \textit{FL with MR}  &   0.95    &  \textbf{0.92}    &   \textbf{0.93}  &   \textbf{0.91} \\
                                    \midrule
                                    
\multirow{5}{*}{\textit{All subsystems}}& \textit{FL}      &   \textbf{1.00}    &  0.36    &   0.53   &   0.50 \\
                                    & \textit{UN}           &   0.19    &  \textbf{\textcolor{red}{0.07}}     &  \textbf{\textcolor{red}{0.10}}   &  \textbf{\textcolor{red}{0.05}} \\
                                    & \textit{PM}           &   \textbf{\textcolor{red}{0.09}}            &  0.35    &   0.14   &   0.08 \\
                                    & \textit{MR}           &   0.87    &  0.82    &   0.85 &  0.77 \\
                                    & \textit{FL with MR}  &   0.88    &  \textbf{0.93}    &   \textbf{0.91}  &  \textbf{0.85} \\
                                    \bottomrule                                 
\end{tabular}
\end{table*}

To provide context for the evaluation, we compared the proposed approach (\textbf{Monitoring Rules - MR}) against three baseline approaches:
\begin{itemize}
    \item \textbf{\textit{OpenStack Failure Logging Mechanisms}} (FL): The OpenStack built-in failure logging mechanisms, which notify the tenants via API errors if the system is not able to serve requests;
    
    \item \textbf{\textit{Non--session-aware approach using unseen n-grams}} (UN): A non--session-aware approach based on unseen \textit{n}-grams~(similar to \cite{an2017behavioral}), where the \textit{n}-gram represents a contiguous sequence of $n$ events within a trace. The approach learns a \textit{normal dictionary} consisting of all occurring \textit{n}-grams from the fault-free traces. During the detection, the approach notifies a failure when a new \textit{n}-gram that is not in the dictionary occurs;
    
    \item \textbf{\textit{Non--session-aware approach using a probabilistic model}} (PM): A non--session-aware probabilistic approach using \textit{Variable-order Markov Models}~\cite{begleiter2004prediction}. The approach is trained on the sequence of the events that occurred in the system under fault-free conditions. The approach notifies a failure whenever the probability of an event occurring after a sequence of events is lower than a fixed threshold $\epsilon_{PM}$. 
\end{itemize}

Moreover, to estimate the improvement obtained by implementing an external monitoring solution to support OpenStack, we evaluated also the performance of the \textbf{\textit{OpenStack failure logging mechanisms combined with the MR approach}} (FL with MR).
All the baseline approaches can perform failure detection at run-time. To perform a fair comparison with the MR approach and highlight the challenges of detecting failures without using IDs, we limited the evaluation to non-session-aware approaches. 
Moreover, we did not include more complex approaches due to the limited number of fault-free traces (i.e., $50$) used for training. In fact, run-time monitoring solutions based on deep learning neural networks require a huge amount of data to detect anomalies in large-scale cloud computing platforms~\cite{islam2021anomaly}.


For each OpenStack subsystem targeted during the fault injection campaign, \tablename{}~\ref{rv:tab:evaluation} shows the results obtained in terms of precision, recall, $F_1$ score, and accuracy. To perform a fair evaluation, we conducted a sensitivity analysis for the UN and PM approaches by varying the number of \textit{n}-grams between 1 and 5~\cite{whalen2014model,shin2020comparison} and the threshold $\epsilon_{PM}$ between $0\%$ and $100\%$. The table shows only the results obtained with the best configuration of the baseline approaches (for UN, set $n=3$; for PM, set $\epsilon_{PM} = 1\%$).

The results highlight that the FL approach provides perfect precision over all the subsystems because the logging mechanisms of the system are not affected by false positive cases (i.e., an API error always implies a system failure). However, the recall provided by this approach is dramatically low ($0.36$ over all the subsystems) since OpenStack is not able to timely detect and notify the failures in many experiments due to missing logging mechanisms, as shown in previous studies~\cite{cotroneo2019bad,marques2022injecting}. 

Different from the system's logging mechanisms, the precision provided by the MR approach is not perfect because, in some cases, the approach wrongly identified a failure due to the non-determinism of the system. Nevertheless, the precision achieved is still very close to the one provided by the FL approach. The considerations on the false-negative cases, instead, are way different. 
Indeed, the results highlight how the MR approach can effectively bring a substantial improvement in the recall values over all the subsystems since the monitoring rules are able to catch the out-of-order or missing events in most of the cases (recall equal to $0.82$). The approach provides worse performance only for the Neutron subsystem. We attribute this to the asynchronous nature of the network service, causing either a missing detection or a missing activation of the rules. 

The results also show that the precision and the recall provided by both the other two non--session-aware baseline approaches, i.e., the approach based on the observation of the unseen $n$-grams (UN approach) and the probabilistic model approach (PM), are not comparable to one of the other approaches over all the metrics, regardless of the target subsystem. 
Indeed, both approaches are not able to fully model the behavior of a multi-tenant and concurrent system without discerning the calls executed by different tenants (using session IDs), given the massive number of events per trace.

The $F_1$ score and the accuracy allow us to compare the approaches both in terms of false positives and false negatives, and thus provides a comprehensive evaluation of the approaches. The metrics have similar values over all the subsystems and suggest that, for the fault injection experiments, the MR approach massively improves the performance obtained with the plain OpenStack logging mechanisms (85\% vs 53\% for the $F_1$, 77\% vs 50\% for the accuracy). In particular, the proposed approach achieves an $F_1$ score and accuracy higher for Nova and Cinder subsystems, while the performance is again worse for the Neutron subsystem. 
The UN and the PM approaches are very far from the performance of the MR approach as they provide a $F_1$ score equal to $0.10$ and $0.14$, respectively. The accuracy is even worse ($0.05$ for UN, $0.08$ for PM) since both approaches are not able to provide true negative cases (i.e., they always identify a failure also when the system is not failed at all). This further emphasizes the results obtained by the MR approach without using any session IDs.

Finally, when the monitoring rules are used in combination with the OpenStack logging mechanisms (FL with MR approach), we can notice that, although the rules slightly impact the precision of the system by wrongly notifying a failure due to the non-determinism of the system, they massively reduced the false-negative cases, overall the subsystems. Even for the Neutron subsystem, when the recall for the proposed approach is lower than OpenStack logging mechanisms, the FL with MR approach takes advantage of the monitoring rules since they help OpenStack to notify failures not detected by the logs of the system.

\subsection{Detection Latency}

\begin{figure}
    \centering
    \includegraphics[width=1\columnwidth]{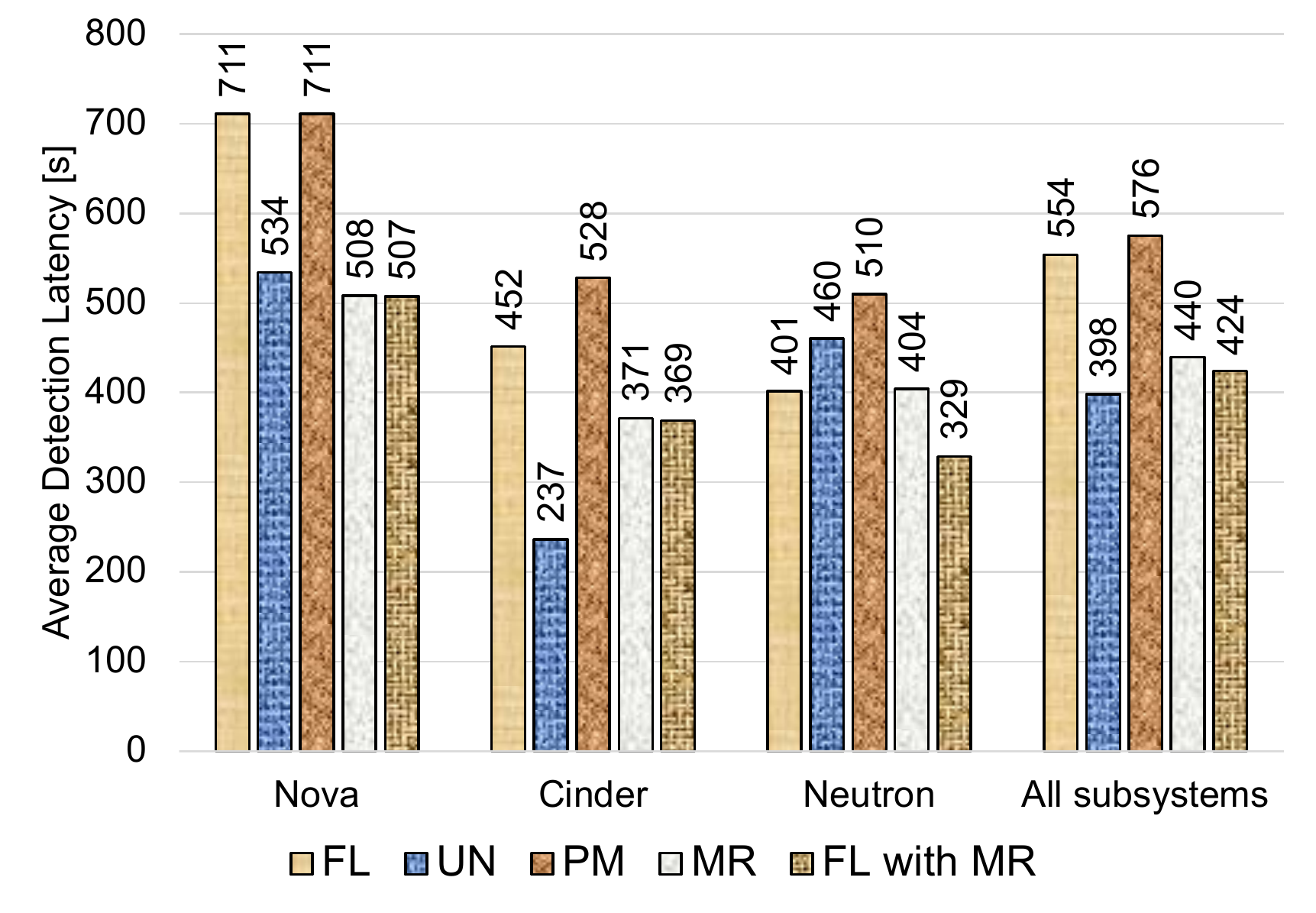}
    \caption{Average detection latency of the different approaches. Time window $\Delta T = 35 $ s (MR approach).}
    \label{rv:fig:detection_latency}
\end{figure}

To provide a more comprehensive evaluation, we also analyzed the promptness of the targeted approaches in the identification of failures. Ideally, a failure should be identified as soon as the system experiences it to quickly restore services and thus increase the reliability of the system.
Therefore, we performed a comparison in terms of \textit{failure detection latency} by computing the time difference between the failure detection time ($t_{fail}$) and a \textit{common starting time}. Given the high non-determinism of the target system, it is not trivial to identify a reliable fault activation time \cite{avivzienis2004dependability,avizienis2004basic}, especially in a multi-tenant scenario. To overcome this issue, we used the starting time of the workload ($t_{start}$) as the \textit{common starting time}, thus the failure detection latency is equal to $t_{fail} - t_{start}$.
Because the approaches are compared over the same experiments and under the same conditions, a shorter failure detection latency indicates the ability to quickly detect failures. To perform a fair comparison, we did not include the false positive cases in this analysis.

\figurename{}~\ref{rv:fig:detection_latency} shows the average failure detection latency (in seconds) provided by the approaches over all fault injection experiments. 
The figure shows that the MR approach provides a notably lower failure detection latency when compared to the FL approach for Nova and Cinder subsystems, and a comparable detection latency for the Neutron subsystem. Overall the fault injection experiments, the average detection latency of MR is $\sim114$ seconds lower than the average detection latency of FL. The failure detection latency of the FL with MR is very close to the MR approach and thus proves that the contribution of the monitoring rules is crucial for the prompt detection of failures at run-time. Also for the Neutron subsystem, where the MR approach showed the worst performance due to the asynchronous nature of the network operations, the FL with MR approach notably decreases the average failure detection latency with respect to the FL approach ($\sim77$ seconds). 
Finally, it is worth noticing that the UN approach, which provides the worst performance in terms of accuracy in detecting the failures, shows the lowest failure detection latency across all subsystems, as it raises an exception every time an unseen sequence of n-grams is observed. The probabilistic approach (PM), instead, is not able to provide timely detection of the failures due to the threshold value, which is set very low (i.e., $1\%$) to limit as many as possible false positive cases. Indeed, since the approach identifies a failure only when the occurrence of an event is very unlikely, then the average time to identify the failure increased.

\subsection{Sensitivity Analysis}

\begin{table}[t]
\centering
\caption{Sensitivity analysis of the time window $\Delta T$ across all subsystems. The best performance is \textbf{bold}. Worst performance is \textcolor{red}{\textbf{red/bold}}}
\label{rv:tab:sensitivity}
\small
\begin{tabular}{>{\centering\arraybackslash}m{3cm} | >{\centering\arraybackslash}m{0.75cm} 
>{\centering\arraybackslash}m{0.75cm} 
>{\centering\arraybackslash}m{0.75cm}  >{\centering\arraybackslash}m{0.75cm}}
\toprule
& \multicolumn{4}{c}{$\Delta T$ (seconds)}\\
\textbf{Metric} & 5 & 20 & 35 & 50\\
\midrule
\textit{Precision} & \textbf{\textcolor{red}{0.73}} & 0.83 & 0.87 & \textbf{0.89}\\
\textit{Recall} & \textbf{\textcolor{red}{0.77}} & \textbf{0.82} & \textbf{0.82} & 0.81\\ 
\textit{$F_1$ score} & \textbf{\textcolor{red}{0.75}} & 0.83 & \textbf{0.85} & \textbf{0.85}\\ 
\textit{Accuracy} & \textbf{\textcolor{red}{0.60}} & 0.74 & \textbf{0.77} & \textbf{0.77}\\
\midrule
\textit{Detection Latency} (seconds) & \textbf{366.26} & 423.63 &  439.80 & \textbf{\textcolor{red}{457.81}}\\
\bottomrule                                 
\end{tabular}
\end{table}

In the previous analysis, we adopted a conservative value for the time window by setting it equal to the maximum time needed by OpenStack to serve any tenant's request in our setup.
Since the choice of the time window influences the length of the patterns, we performed a sensitivity analysis. 
\tablename{}~\ref{rv:tab:sensitivity} shows the results of the MR approach by setting the time window $\Delta T$ equal to $5$, $20$, $35$, and $50$ seconds. 

Unsurprisingly, we found that the performance of the approach improves by increasing the time window since a shorter $\Delta T$ increases the number of false positives and limits the true negatives. As matter of fact, we found that when $\Delta T$ is equal to $5$ seconds, the approach provides a \textit{false positive rate} equal to $1$. 
The table also shows that, although the increment of the time window implies an improvement of the precision, the recall saturates when $\Delta T$ is equal to $20$ seconds, and slightly decreases when the time window is set to $50$ seconds since a pattern too temporally long is affected by false-negative cases. Therefore, both the $F_1$ score and the accuracy of the MR approach saturates when $\Delta T$ is higher than $35$ seconds.
Moreover, as expected, the table shows that a larger time window implies an increment in failure detection latency. Since the choice of the time window should be a valid trade-off between the ability to detect failures and the failure detection latency, a time window equal to $35$ seconds (or also $20$ seconds) is considered a more proper choice.

\subsection{Computational Cost}
We performed the analysis (on the same system used for the experiments) of the computational cost required to derive the monitoring rules from fault-free traces.
The computational cost includes the time needed to parse the logs, filter events, and run the algorithm to find the patterns. We found that the overall time needed to simultaneously analyze $50$ different fault-free execution traces (which contain $\sim 120K$ rows) is lower than $70$ seconds (i.e., less than $1.5$ seconds per trace, on average). The computational cost increases linearly with the number of traces.

\section{Threats to Validity}
\label{rv:sec:threats}

\noindent
\textbf{Case Study.} OpenStack is one of the most widely deployed open-source cloud software in the world and represents an important case study. 
The execution of fault-injection experiments on OpenStack is non-trivial because, to guarantee independence among the experiments, we restart all the services, restore the database, clean up the files, execute the workload, etc (every experiment lasts 1 hour on average). Therefore, the application of the approach and the execution of the fault-injection experiments also on other cloud computing platforms is very time-consuming.
To mitigate this threat to validity, we targeted three systems from the OpenStack umbrella project (i.e., Nova, Neutron, Cinder), which are large and diverse enough to get interesting insights into the application of the proposed approach across different independent systems. This diversity is reflected by differences both in terms of project-specific patterns (programming idioms, API conventions), and different events exchanged in the systems (number, type, and non-determinism).

\vspace{1pt}
\noindent
\textbf{Baseline approaches.} To perform a fair evaluation, we compared the results of the MR approach with two non--session-aware approaches. The choice of the baseline approaches also depended on the public availability of open-source code to reproduce the experiments and the portability of the approaches on a different system such as OpenStack.
We adopted an n-gram-based solution and a probabilistic model approach because sequence analysis approaches are widely used in practice. As matter of fact, there are several research studies adopting n-grams (or similar algorithms) and probabilistic models to perform run-time detection of anomalies up to now~\cite{khreich2017anomaly,ariff2021ensemble,brown2022online,cailliau2019runtime,carreon2021probabilistic,bartolo2021towards}. 
Moreover, we did not include more complex approaches, such as neural networks-based approaches, since they require massive data for training~\cite{islam2021anomaly,huch2018machine,girish2021anomaly}. Indeed, the collection of the system executions is an essential concern in our context as developers have a limited time budget to spend for fault-injection testing. Since executions can take several hours in commercial-grade systems, we need to limit the number of fault-free executions (as matter of fact, we collected only $50$ executions).

\section{Conclusion}
\label{rv:sec:conclusion}
In this work, we proposed an approach to run-time verification in cloud computing systems that derives a set of monitoring rules from the fault-free executions of the system. The rules are then synthesized in a monitor solution by using a specification language.
We applied the approach in the OpenStack cloud computing platform, where we evaluated the ability of the monitoring rules in detecting failures in a campaign of fault injection experiments with a multi-tenant workload. 
Our experiments showed that the approach achieves better performance, in terms of $F_1$ score, when compared to the OpenStack logging mechanisms and two non--session-aware run-time verification approaches, and significantly decreases the time to detect the failure at run-time. 
The approach, when used in combination with the failure logging mechanisms of the system, provides an $F_1$ score higher than $90\%$ and accuracy of $85\%$, improving the fault tolerance mechanisms of the system.

The MR approach can be applied in different OpenStack configurations as it does not depend on any configuration parameter, except the size of the time window (i.e., the time duration of the patterns). In our experiments, we made a conservative choice by setting the size equal to the maximum time taken by OpenStack to serve a request in fault-free conditions (35 seconds, in our testbed). The sensitivity analysis of the time window showed that this conservative choice provides the best performance in terms of $F_1$ score and accuracy. Although the maximum time to serve the requests depends on both the workload, i.e., the operations performed during the experiments, and the hardware where OpenStack is deployed (high hardware requirements imply a shorter time to serve the requests), it can be easily computed by running the workload in fault-free conditions (e.g., by using the logs of the workload).



\section*{Acknowledgements}
This work has been partially supported by the University of Naples Federico II in the frame of the Programme F.R.A., project id OSTAGE. 
We are grateful to our former students Nicola Apa and Roberto Scarpati for their help in the early stage of this work.

\bibliographystyle{elsarticle-num} 
\bibliography{biblio,website}

\end{document}